\documentclass[twocolumn,aps, prd, amsmath, floats,floatfix, superscriptaddress, nofootinbib, showkeys]{revtex4}

\usepackage{amssymb}
\usepackage{caption}
\usepackage{amsmath}
\usepackage{verbatim}
\usepackage{mathrsfs}
\usepackage{amsfonts}
\usepackage{latexsym}
\usepackage{epsfig}
\usepackage{color}
\usepackage{hyperref}

\usepackage{units}
\usepackage{envmath}
\usepackage{natbib}
\usepackage{ctable}
\usepackage{subfig}
\usepackage{graphicx}
\begin{document}
\definecolor{orange}{rgb}{0.9,0.45,0}
\def\CovDev{D}
\def\Res{{\mathcal R}}
\def\Gammaflat{\hat \Gamma}
\def\metricflat{\hat \gamma}
\def\Dflat{\hat {\mathcal D}}
\def\part_n{\partial_\perp}
%
\def\Lie{\mathcal{L}}
\def\A{\mathcal{X}}
\def\Aphi{\A_{\phi}}
\def\hAphi{\hat{\A}_{\phi}}
\def\E{\mathcal{E}}
\def\Ham{\mathcal{H}}
\def\M{\mathcal{M}}
\def\R{\mathcal{R}}
\def\p{\partial}
\def\hg{\hat{\gamma}}
\def\hA{\hat{A}}
\def\hD{\hat{D}}
\def\hE{\hat{E}}
\def\hR{\hat{R}}
\def\hcA{\hat{\mathcal{A}}}
\def\hDelt{\hat{\triangle}}
\def\na{\nabla}
\def\dif{{\rm{d}}}
\def\non{\nonumber}
\newcommand{\erf}{\textrm{erf}}
%
\renewcommand{\t}{\times}
\long\def\symbolfootnote[#1]#2{\begingroup%
\def\thefootnote{\fnsymbol{footnote}}\footnote[#1]{#2}\endgroup}
\title{Modeling the Central Supermassive Black
Holes Mass of Quasars via LSTM Approach}

\author{Seyed Sajad Tabasi} 
\email{sstabasi98@gmail.com}
\affiliation{Department of Physics, Sharif University of Technology,
P. O. Box 11155-9161, Tehran, Iran}
	\affiliation{PDAT Laboratory, Department of Physics, K. N. Toosi University of Technology, P.O. Box 15875-4416, Tehran, Iran}
\author{Reyhaneh Vojoudi Salmani}
\email{r.s.vojoudi@gmail.com}
\affiliation{Department of Physics, K. N. Toosi University of Technology, P. O. Box 15875-4416, Tehran, Iran}
\affiliation{PDAT Laboratory, Department of Physics, K. N. Toosi University of Technology, P.O. Box 15875-4416, Tehran, Iran}
\author{Pouriya Khaliliyan}
\email{pouriya@email.kntu.ac.ir} 
\affiliation{Department of Physics, K. N. Toosi University of Technology, P. O. Box 15875-4416, Tehran, Iran}
\affiliation{PDAT Laboratory, Department of Physics, K. N. Toosi University of Technology, P.O. Box 15875-4416, Tehran, Iran}
\author{Javad T. Firouzjaee}
\email{firouzjaee@kntu.ac.ir}
\affiliation{Department of Physics, K. N. Toosi University of Technology, P. O. Box 15875-4416, Tehran, Iran}
\affiliation{PDAT Laboratory, Department of Physics, K. N. Toosi University of Technology, P.O. Box 15875-4416, Tehran, Iran}
\affiliation{ School of Physics, Institute for Research in Fundamental Sciences (IPM), P. O. Box 19395-5531, Tehran, Iran } 

\begin{abstract}
One of the fundamental questions about quasars is related to their central supermassive black holes. The reason for the existence of these black holes with such a huge mass is still unclear and various models have been proposed to explain them. However, there is still no comprehensive explanation that is accepted by the community. The only thing we are sure of is that these black holes were not created by the collapse of giant stars, nor by the accretion of matter around them. Moreover, another important question is the mass distribution of these black holes over time. Observations have shown that if we go back through redshift, we see black holes with more masses, and after passing the peak of star formation redshift, this procedure decreases. Nevertheless, the exact redshift of this peak is still controversial. In this paper, with the help of deep learning and the LSTM algorithm, we tried to find a suitable model for the mass of central black holes of quasars over time by considering QuasarNET data. Our model was built with these data reported from redshift 3 to 7 and for two redshift intervals 0 to 3 and 7 to 10, it predicted the mass of the quasar's central supermassive black holes. We have also tested our model for the specified intervals with observed data from central black holes and discussed the results.
 
\end{abstract}

\keywords{Quasars, Supermassive Black Holes, Sloan Digital Sky Survey, QuasarNET Data Set, Deep Learning, and LSTM Model}

\maketitle
\section{Introduction}

In recent years, the study of the high-redshift($z>6$) quasars was a direct probe to explore the Universe at the age less than 1 Gyr after the Big Bang. These early forming quasars are essential to studying the early growth of supermassive black holes (SMBHs) \cite{inayoshi2019assembly}. 

By detecting the reverberation between the variations of broad emission lines and the continuum we can determine SMBHs mass in quasars \cite{blandford1982reverberation}. Until now, the time lag of H$\beta$ emission lines has been confirmed and measured only in $\sim  $100 quasars \cite{du2019radius}.

The continuum and line emission from luminous quasars which are one of the most luminous objects, over a large wavelength range can be characterized by several leading parts. The broad emission line region \cite{antonucci1993unified} 
the optical-to-ultraviolet continuum emission, which is explained by a standard accretion disk extending down to the innermost stable circular orbit \cite{shields1978thermal}, 
X-ray emission with a power-law spectrum produced by inverse Compton scattering of photons from the accretion disk of relativistic electrons in the hot corona \cite{svensson1994black}, and a soft X-ray excess \cite{arnaud1985exosat}. Spectroscopic observations from optical to near-infrared of these
quasars suggest that such SMBHs are already established when the universe is only $ 700 Myr $ old \cite{yang2020poniua}.

To explain the existence of these SMBHs, many theoretical models have been proposed like using primordial density seeds \cite{wise2019formation,kroupa2020very,bernal2018signatures} and appealing a super-Eddington accretion process \cite{volonteri2015case}.

To utilize the spectroscopic observational data in physical studies, we need an exact classification and redshift determination of astrophysical objects. Along the way, the Sloan Digital Sky Survey Catalogue 16th Data Release Quasar Only(SDSS-DR16Q) \cite{Lyke2020}, consists of two files, being the quasar-only main catalog of 750414 entries which contains sooner visually confirmed quasars  SDSS-I/II/III, and a 1440615-row “superset” of SDSS-IV/eBOSS quasar object classifications.

The DR16Q catalogs present multiple redshifts per object that are available, including the neural automated QuasarNET \cite{Busca2018} redshift which is claimed $>99\%$ efficiency and $>99\%$ accuracy, that rests on garnering deeper insights into this triumvirate connection by co-locating and analyzing observational data and simulated data. Meanwhile, the enormous increase in computing power over the last decades has allowed the application of acquired statistical methods in the analysis of big and complex data sets.

Using previously-fed data has brought huge opportunities for astronomers to develop intelligent tools and interfaces, utilizing pipeline classifiers, machine learning(ML), and deep learning(DL) methods, to deal with data sets and extract novel information with possible predictions and estimate the relevant confidence which the behavior new data will have.

In astronomy and astrophysics, ML \cite{ball2010data,baron2019machine} and  DL \cite{allen2019deep, meher2021deep} have been used in a broad range of subjects(e.g. quasars and other types of sources), such as redshift determination \cite{nakoneczny2021photometric,wenzl2021random}, morphological classification and references therein \cite{vardoulaki2021fr, burhanudin2021light},
source selection and classification \cite{parkinson2016classification,xiao2020efficient,wang2022j}, image and spectral reconstruction \cite{li2021ai}, and more.

ML methods for obtaining redshift estimation for quasars are becoming progressively crucial in the epoch of rich data astronomy. Redshift measurements of quasars are important as they can enable quasar population studies, and provide insight into the star formation rate(SFR), the luminosity function(LF), and the density rate evolution \cite{dainotti2021predicting}.

In this work, we have used DL to model the mass of quasars' central SMBH as a function of their redshift. Firstly, Sec. II is dedicated to the available observational data and evidence on quasars. The estimation of a quasar's central SMBH mass is discussed in detail in Sec. III. Furthermore, in Sec. IV, the mass evolution of these black holes(BHs) is investigated. Sec. V is the comparison between two newborn research platforms, QuasarNET and FNET, and the reasons behind using QuasarNET for our model are explained. Additionally, we use two correction methods which are explained in Sec. VI. A detailed explanation of our DL model can be found in Sec. VII to X. In Sec. VII we introduce Long short-term memory(LSTM) which is the recurrent neural network(RNN) that we build our model based on. We explain the chosen optimization function and its validation loss in Sec. VIII which is shown in multiple figures. Sec. IX presents the topology design of our model and finally, the comparison of the model predictions with other data sets is discussed in Sec. X.\\
\maketitle
\section{Observational evidence and data}

The most comprehensive observed quasi-stellar objects(QSOs) spectra to date are cataloged in the SDSS-IV. SDSS has been operative since 2000 and catalogs of quasars have been produced and made available since 2002. In addition to producing images, it performs spectroscopic surveys across a large area of the sky. We can get about one million galaxies and 10,000 quasars spectra from the survey images of the sky, which are obtained by a 2.5m telescope equipped with a large format mosaic Charge-coupled device(CCD) camera, and two digital spectrographs. As part of its calibration, the SDSS uses observations of the US Naval Observatory's 1m telescope to calibrate its photometry, and an array of astrometric CCDs control its astrometry \cite{York2000}.

The SDSS provides data necessary to study the large-scale structure of the universe. As far as the observatory's limit allows, the imaging survey should detect $\sim5\times10^7$ galaxies, $\sim10^6$ quasars, and $\sim8\times10^7$ stars. By using photometric redshifts and angular correlation functions, these photometric data allow studies of large-scale structures that go beyond spectroscopic analysis. Quasars can provide information on the structure at even larger scales \cite{York2000}.

The SDSS-DR16Q contains 750,414 quasars, with the automated redshift range $1\leq z \leq7.1$. The number of sources reaches its maximum around $z \approx 2.5$ and at earlier epochs i.e. higher redshifts, they are comparatively rare \cite{Rastegarnia2022}.
However, there is a problem with the SDSS-DR16Q catalog. It contains non quasar sources due to pipeline classification errors and incorrect redshift estimations \cite{Lyke2020}. For example, in a search for undeclared quasars, the SDSS-DR16Q main quasars are found to contain 81 entries that are not quasars. It must therefore be noted that the pipeline catalog is not an adequate training samples for quasars because many objects with $z\geq6$ as well as significant fractions of these objects at $z\geq4$, may not be quasars or not quasars at the given redshifts due to incorrect pipeline classifications \cite{Rastegarnia2022}.

\maketitle
\section{Mass estimation of quasars' central SMBH}

In terms of fundamental parameters of quasars, one can mention the central SMBH mass and structure, along with the ratio of the accretion rate to the Eddington accretion rate \cite{Chen2005}.

The central SMBH mass can be measured via the gas or stellar dynamics \cite{Chen2005} from optical or ultraviolet(UV) spectroscopy using empirical relations \cite{Vestergaard2002}. 
The broad emission line region(BLR) probably provides the best probe of these characteristics \cite {Wandel1999}. The size of BLRs can be determined by reverberation mapping(RM) \cite{Rodriguez1997}, which is a measuring technique in astrophysics. RM provides
invaluable information about the kinematic and ionization distribution of the gas using the time lag between emission line and continuum variations \cite{Wandel1999}. 

Assuming that gravity dominates the dynamics of the BLR and the virial relationship between time lag
and line width exists, the BH mass can be estimated as \cite{Bentz2009}
\begin{equation}
\label{eqn:eq1}
     M_{BH}= \frac{f c \tau v^2}{G},
\end{equation}

where $\tau$ is the mean time delay for the region of interest, $v$ is the velocity of the gas in that region, $c$ is the speed of light, $G$ is the gravitational constant, and $f$ is a scaling factor of order unity that depends on the detailed geometry and kinematics of the line-emitting region.

The worth mentioning point is that the virial relationship claims a virialized system with individual clouds moving in their Keplerian orbits. This leads to the proportionality of mean cloud velocity and emissivity radius \cite{Netzer2013}
\begin{equation}
\label{eqn:eq2}
    v \propto{r_{BLR}}^{\frac{1}{2}},
\end{equation}

where $r_{BLR}$ is the emissivity radius.

In the absence of RM, the quasar continuum luminosity is sufficient to estimate the BLR. With RM estimations, the best-fitting $R_{BLR}-\lambda L_\lambda$ relations were derived for quasars at monochromatic luminosity in both 3000 and 5100 $\AA$ rest-frames as follows \cite{McLure2002}

\begin{equation}
\label{eqn:eq3}
    R_{BLR} = (18.5 \pm 6.6){[\lambda L_{3000} / 10^{37} W]}^{(0.32 \pm 0.14)},
\end{equation}

\begin{equation}
\label{eqn:eq4}
    R_{BLR} = (26.4 \pm 4.4){[\lambda L_{5100} / 10^{37} W]}^{(0.61 \pm 0.10)}.
\end{equation}

Here, $L$ is the luminosity measured at a
wavelength $\lambda$.
In Eq. \ref{eqn:eq1},
an intrinsic Keplerian velocity of a broad-line gas is related to the  full width at half maximum (FWHM) of a chosen broad emission line by the geometric factor $f$ as 

\begin{equation}
\label{eqn:eq5}
    V_{BLR} = f\times FWHM,
    \end{equation}
    
 In other words, it is the width of a spectrum curve measured between those points on the y-axis which are half of the maximum amplitude. 

As the geometry of the BLR in radio-quiet quasars is currently unknown, it is generally agreed that $f = \sqrt{3/2}$, which is appropriate for randomly oriented orbits of the BLR gas. However, FWHM measurements for broad emission lines in radio-loud quasars indicate a disc-like geometry \cite{Wills1986}.
Given the similarity between the optical emission-line spectra of radio-loud and radio-quiet quasars, it is not unreasonable to consider the possibility that BLRs of radio-quiet quasars that dominate the SDSS data can follow the same equation as well \cite{McLure2004}

\begin{equation}
\label{eqn:eq6}
    V_{BLR} = \frac{FWHM}{(2 \sin i)}.
\end{equation}

Here, $i$ represents the angle between the line of sight and the axis of the disc.

Our virial BH mass estimators are derived by substituting the calibrations of the $R_{BLR}–\lambda L_\lambda$ relations into Eq. \ref{eqn:eq1} and determining $V_{BLR}$ using $Mg II$ or $H\beta$ \cite{McLure2002}.

 Based on the  $L_{5100}$ which is the monochromatic luminosity at rest-frame 5100 $\AA$ and the $H\beta$ line, a more specific expression to calculate the mass of a BH can be written as \cite{Koss2017}
 
 \begin{eqnarray}
 \label{eqn:eq7}
    M_{BH}(H\beta) &=& 1.05 \times 10^8 (\frac{L_{5100}}{10^{46}ergs^{-1}})^{0.65} \\
     \nonumber
    &\times&[\frac{FWHM({H\beta)}}{10^3 kms^{-1}}]^2 M_{\odot},
\end{eqnarray}

where $M_{BH}(H\beta)$ represents BH mass by considering $H\beta$ line, $FWHM({H\beta})$ is the full width at half maximum of $H\beta$ line, and $M_\odot$ is the solar mass.

Large spectroscopic surveys like the SDSS observe both broad $H\beta$ and $Mg II$ lines. Therefore, one can be calibrated against the other and based on $L_{3000}$ and $Mg II \lambda 2798$ line width, a similar expression can be derived as \cite{Netzer2013}

\begin{eqnarray}
\label{eqn:eq8}
    M_{BH}(MgII\lambda2798) &=& 8.9 \times 10^7 (\frac{L_{3000}}{10^{46}ergs^{-1}})^{0.58}\\
    \nonumber
    &\times&[\frac{FWHM(MgII\lambda{2798})}{10^3 kms^{-1}}]^2 M_{\odot},
\end{eqnarray}

where $M_{BH}(MgII \lambda2798)$ represents BH mass by considering $H\beta$ line, and $FWHM({MgII \lambda2798})$ is the full width at half maximum of $MgII$ line.

Based on empirical estimation of $f\simeq 1.1$ for the $H\beta$ line, we can now write more specific expressions to calculate $M_{BH}$ for several emission lines like $MgII$ as follows  \cite{Koss2017} 

\begin{eqnarray}
\label{eqn:eq10}
\frac{M_{BH}}{M_\odot}  &=& 4.7 {(\frac{\lambda L_{5100}}{10^{37} W})}^{0.61} {[\frac{FWHM (H\beta)}{kms^{-1}}]}^2, \\
\label{eqn:eq9}
\frac{M_{BH}}{M_\odot}  &=& 3.2 {(\frac{\lambda L_{3200}}{10^{37} W})}^{0.62} {[\frac{FWHM (Mg II)}{kms^{-1}}]}^2.
\end{eqnarray}

Besides, it is well-known that the relationship between stellar velocity dispersion and BH mass can be written as \cite{Koss2017}

\begin{equation}
\label{eqn:eq11}
    log(\frac{M_{BH}}{M_{\odot}}) = 4.38 \times log (\frac{\sigma_{*}}{200kms^{-1}}) +8.49,
\end{equation}

where $\sigma_{*}$ is the stellar velocity dispersion.

Furthermore, to estimate the mass of a BH, observations in the local universe reveal the existence of a correlation between the central SMBH mass and the bulge of the host galaxies \cite{Schutte2019}.

\begin{equation}
\label{eqn:eq12}
    log (\frac{M_{BH}}{M_{\odot}}) = \alpha + \beta log 
    (\frac{M_{Bulge,*}}{10^{11} M_{\odot}}),
\end{equation}

where $M_{Bulge,*}$ is the bulge stellar mass and the best-fit of $\alpha$ and $\beta$ should be

\begin{equation}
\label{eqn:eq13}
       \alpha = 7.93 \pm 0.061 ; \,\beta = 1.15 \pm 0.075.
\end{equation}

\section{Mass evolution of quasars' central SMBH}

As studying the cosmic history of compact cosmological objects is so crucial to track the history line of the universe in a much bigger structure, we are so curious about the evolution of SMBHs mass. In the presence of a SMBH, there are obvious links between the physical properties and those of its host. Due to high redshifts that many quasars have, they are ideal to be studied to recognize BH evolution through time back to the early universe \cite{Willott2010}.

According to the modelling of spectra from the SDSS first data release, the virial mass of BHs for 12698 quasars in the redshift interval $0.1 \leq z \leq 2.1$ is estimated. There is entirely consistent evidence to suggest that the BH mass of SDSS quasars lies in $10^7 M_{\odot} \leq M_{BH} \leq 3 \times 10^9 M_{\odot}$. The local BH mass function for early-type galaxies using the $M_{BH}-\sigma$ and $M_{BH}- L_{bulge}$ correlations(Eq. \ref{eqn:eq11} and Eq. \ref{eqn:eq12}) are also estimated.
In addition, by comparing the number density of active BHs at $z \approx 2$ with the local mass density of inactive ones, a lower limit is set on the lifetime of quasars, which confirms that the bulk of BHs with mass $\geq 10^{8.5} M_{\odot}$ are situated in place by $z \approx 2$ \cite{McLure2004}.

There are several different ideas on the central SMBH mass evolution through time in literature. Based on the effective flux limit along with the role of the quasar continuum luminosity, most studies agree that the SMBH mass increases as a function of redshift, namely most low mass SMBHs can be found in the late universe(e.g. stepping down from $\approx10^9 M_{\odot}$ at $z\approx
2.0$ to $\approx10^8 M_{\odot}$ at $z \approx 0.2$). Considering Eq. \ref{eqn:eq10} and Eq. \ref{eqn:eq9}, redshift does not alter the mean FWHM and it can be roughly considered to be constant. Therefore, the mean virial mass of the SMBH should be increased as $ [L_{\lambda}]^{0.6}$ \cite{McLure2004}.

Quasars undergo important cosmic evolution according to optical, X-ray, and bolometric LFs. Interestingly, based on predictions of  \cite{Fanidakis2012} using an extended version of the galaxy formation model,  GALFORM code, quasars evolution will be influenced by different physical processes such as the accretion mode and the obscuration prescription. Observational data have also reported similar trends \cite{Hopkins2007}.

Furthermore, SMBHs grow exponentially during a period in which accretion governs their mass evolution. When $z\gtrsim5$, the growth of a SMBH in a quasar is as follows \cite{Benny2019}

\begin{eqnarray}
\label{eqn:eq115}
     M_{BH}(t) &=& M_{BH}(t_0) e^{t\tau}, \\
     \tau &\simeq& 0.4 Gyr \frac{\eta}{1-\eta} \frac{1}{\mu}, \\
     \mu &\equiv& \frac{L}{L_{Edd}}\times f_{active},
\end{eqnarray} 

where $M_{BH}(t_0)$ is the initial mass of BH i.e. the seed's mass, $\eta$ is the radiative efficiency(see \cite{Benny2017} for reported values of $\eta$ for several objects), $L$ is the luminosity of the quasar, $L_{Edd}$ is the luminosity at Eddington limit, $f_{active}$ is the duty cycle, and $\mu$ is a constant which is determined as a combination of $L/L_{Edd}$ and $f_{active}$. Therefore, it is possible to calculate the growth of the BH easily as

\begin{eqnarray}
\label{eqn:eq1157}
      \log{M_{BH}}(z) &=& \log{M_{BH}}(z_0) \\
      \nonumber
      &+& \log [\exp{(R (\frac{1-\eta}{\eta})z_d}), \\
      \eta &\equiv&\frac{L_{bol}}{ \Dot{M}c^2},\\ 
      z_d &\equiv&  (1+z)^{-3/2} - (1+z_0)^{-3/2}.
\end{eqnarray} 

In above equations, $M_{BH}(z_0)$ is the mass of BHs' seed and $R$ is a constant that is defined as follows

\begin{eqnarray}
R &\equiv& \frac{0.4 Gyr}{\mu},\\
R&=& \begin{cases}
3.79322, \qquad \mu=0.1\\
18.9661, \qquad \mu=0.5\\
37.9322, \qquad \mu=1.0.\\
\end{cases}
\end{eqnarray}

\section{QuasarNET and FNET }

To investigate the mass evolution even more precisely, QuasarNET and FNET are the two available research platforms. Using ML, QuasarNET makes deployment of data-driven modelling techniques possible by combining and co-locating large observational data sets of quasars, the high-redshift luminous population of accreting BHs, at $z\geq3$ alongside simulated data spanning the same cosmic epochs. The main quasar population data source of QuasarNET is NASA Extra-galactic Database(NED) which contains quasars retrieved from several independent optical surveys, principally the magnitude-limited SDSS. There is no comparison between quasars from SDSS and those from other surveys when it comes to spectra and photometry \cite{Natarajan2021}. 

NED contains all quasars in principle, but some are missing because their photometric redshifts were incorrectly assigned. Photometric redshift estimation methods suffer from degeneracy, a well-known limitation of current photometric redshift determination methods \cite{Salvato2019}. QuasarNET fills in the missing sources by analyzing the published catalogues from all surveys. It expands to include additional parameters used to
derive BHs mass, instead of archiving only the reported masses. It contains 136 quasars' features, such as the position, redshift, luminosity, mass, line width, and 
Eddington ratio. 

Two observationally determined functions are used as constraints in theoretical models to describe the assembly history of the BHs population across time: the BH mass function and the Quasar Luminosity Function(QLF). As a statistical measurement of the combined distribution of BHs mass through redshifts, the BH mass function encodes the mass growth history. Similar to the QLF, which reflects their accretion history, the BH mass function is a statistical measurement of the distribution of quasars' luminosities through redshift \cite{Natarajan2021}. 

On the other hand, by using DL, to study quasars in the SDSS-DR16Q of eBOSS on a wide range of signal-to-noise(SNR) ratios, there is a 1-dimensional convolutional neural network(CNN) with a residual neural network(ResNet) structure, named FNet. With its 24 convolutional layers and ResNet structure, which has different kernel sizes of 500, 200, and 15, FNET can use a self-learning process to identify "local" and "global" patterns in the entire sample of spectra \cite{Rastegarnia2022}.

Although FNET seems to be similar to the recently adopted CNN-based redshift estimator and classifier, i.e. QuasarNET \cite{Busca2018}, their hidden layer implementations are distinct.

The redshift estimation in FNET is done based on relating the hidden pattern which lies in flux to a specific redshift, not using any information about emission/absorption lines, while QuasarNET
follows the traditional redshift estimation procedure using the identified emission lines in spectra. This makes FNET to outperform QuasarNET for some complex spectra(insufficient lines, high noise, etc.) by recognizing the global pattern.

Moreover, FNET provides similar accuracy to QuasarNET, but it is applicable for a wider range of SDSS spectra, especially for those missing the clear emission lines exploited by QuasarNET. In more detail, from a statistical point of view, FNET is capable to infer accurate redshifts
even for low SNRs or incomplete spectra. It predicts the redshift of 5,190 quasars with 91.6 \% accuracy, while QuasarNET fails to estimate \cite{Rastegarnia2022}. 

It is important to know that the FNET vs. QuasarNET comes out on top in redshift prediction, but its lack of quasars' central SMBH mass information makes QuasarNET the preferred option for some studies like this work. However, if in the future SMBHs mass will be estimated by using redshifts from FNET approach, our study can be done again to achieve more accurate results.

\section{Flux and volume-limited samples}

Observations are affected by flux as we move to higher redshifts and more distant objects. This is why some objects are not included in data sets. We suppose that they are not even present because their low flux makes them very difficult or in some cases impossible to observe. This will influence the results of any model that is built on a set of objects. To remove this bias, we must first correct the data set.

Two correction methods can be put into use to build a corrected data set and check if the result is solid or if the correction can end up with a huge deviation from the first result. 

Using the friends-of-friends algorithm, quasars can be linked into systems with a specific neighbourhood radius, called linking length($LL$). The size of the group can be determined based on the choice of $LL$ or more generally on its scaling law. $LL$ is parameterized upon a scaling law as \cite{Tago2010}

\begin{equation}
    \frac{LL}{LL_0}=1+a\,\arctan(\frac{z}{z_*}),
\end{equation}
where $a=1.00$, $z_*=0.050$ and $LL_0$ is the value of $LL$ at initial redshift. 

Setting a limit for absolute magnitude is needed for creating volume-limited samples and all less luminous quasars have to be excluded from the data set. Flux-limited samples, on the other hand, are formed from dozens of cylinders containing quasars. Flux-limited samples can be made with both constant and varying $LL$. The constant $LL_0$ is set as \cite{Tago2010} 

\begin{eqnarray}
        LL_0 &=& 250 [kms^{-1}], \\
        LL_0 &=& 0.25 [h^{-1}Mpc].
\end{eqnarray}

Following the extraction of the necessary columns and rejecting duplicate quasars from the data set, there is only one step left, which is verifying if the quasars are within the volume of cylinders generated by the $LL$s.
To do so first we generate a cylinder, then by using the distance between quasars and comparing this distance with the volume of the cylinder, we consider a quasar to be an accepted object if it is located in the cylinder. The distance can be easily obtained from the redshift difference between them in the data set. This algorithm should be repeated as a loop for each quasar.

\begin{figure}
{\includegraphics[width = 3in]{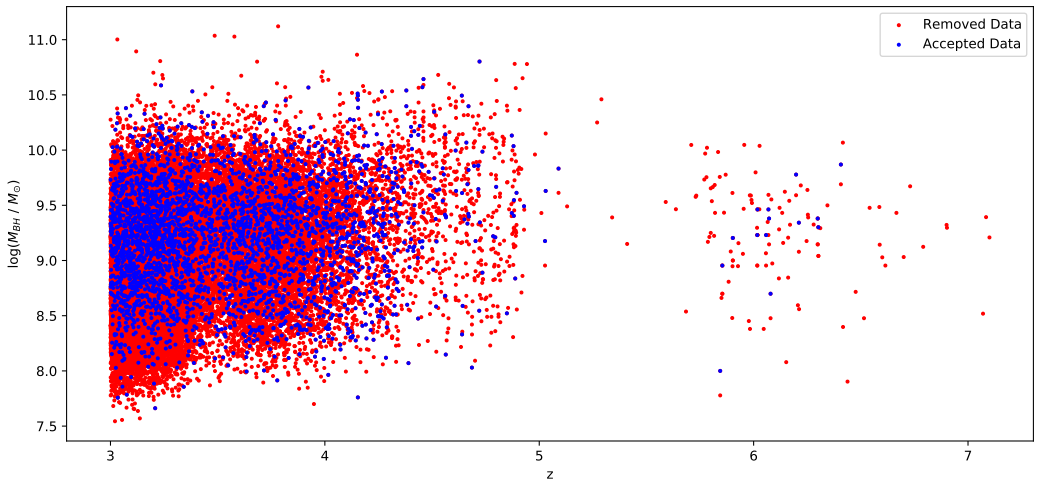}}\\
\caption{ \footnotesize The total number of objects available in the QuasarNET data set is 37648. As a result of data correction methods, 34403 objects were removed (red dots). The accepted data are the final flux and volume-limited samples, made of 3245 Objects(blue dots).}
\label{fig:108}
\end{figure}

As a result of applying the correction methods that are described, we end up with 3246 objects to work with, instead of 37648 objects that are available in QuasarNET. In FIG. \ref{fig:108} accepted and rejected quasars' central SMBH of SDSS-DR16Q in terms of their redshift are illustrated.

\section{Long short-term memory}

LSTM is one of the most powerful RNN that is used in DL and artificial intelligence \cite{Staudemeyer2019}. The RNN is a dynamic system in which there is an internal state at each step of the classification process \cite{Williams1990,Werbos1990}. The circular connections between neurons at the higher and lower layers, as well as the possibility of self-feedback, are responsible for this. These feedback connections enable RNNs to propagate data from earlier events to current processing steps. Thus, RNNs build a memory of time series events.

A standard RNN is not capable of bridging more than 5 to 10 time steps. It is because back-propagated error signals either grow or shrink with every time step \cite{Staudemeyer2019}. As a result, the error typically blows up or disappears over a long period of time \cite{Bengio1994,Hochreiter1996}. When error signals are blown up, the result is oscillating weights, while vanishing errors mean learning takes too long or does not work at all. It is possible to solve the vanishing error problem by using a gradient-based approach known as LSTM \cite{Hochreiter1997,Hochreiter1996,Gers2000,Gers2002}. More than 1,000 discrete time steps can be bridged using LSTM. LSTM uses constant error carousels(CECs), which enforce a constant error flow within special cells. 

Cell accessibility is handled by multiplicative gate units, which learn when to grant access to cells \cite{Staudemeyer2019}. Using a multiplicative input gate unit, memory contents stored in j are protected from irrelevant inputs.  We also introduce a multiplicative output gate unit that protects other units from being perturbed by currently irrelevant memory contents stored in j \cite{Graves2012}. Considering distinct time steps t= 1, 2, etc., an individual step includes forward and backward passes which are the update of all units and calculation of error signals for all weights, respectively. The Input $y^{in}$ and output $y^{out}$ gate activation are computed as \cite{Gers2000}

\begin{eqnarray}
\label{eqn:eq14}
    net_{out_{j}}(t)&=&\sum_{m} \omega_{out_{j}m}y^{m}(t-1),\,y^{out_{j}}(t)\\
    \nonumber
    &=&f_{out_{j}}(net_{out_{j}}(t)),   
\end{eqnarray}

\begin{eqnarray}
\label{eqn:eq15}
    net_{in_{j}}(t)&=&\sum_{m} \omega_{in_{j}m}y^{m}(t-1),\,y^{in_{j}}(t)\\
    \nonumber
    &=&f_{in_{j}}(net_{in_{j}}(t)).   
\end{eqnarray}

Here, $net_{inj}$ and $net_{out}$ are the input and output gate activation, $j$ indices are memory blocks, $\omega_{lm}$ is the weight on the connection from unit $m$ to $l$. Index m ranges over all source units, as specified by the network topology. For gates, $f$ is a logistic sigmoid in the range of $[0, 1]$. 

Furthermore, there are adaptive gates, which learn to reset memory blocks once their contents are out of date and therefore, useless. Like the activation of the other gates(Eq. \ref{eqn:eq14} and Eq. \ref{eqn:eq15}), the forget gate activation $y^{\phi}$ is calculated as

\begin{eqnarray}
\label{eqn:eq16}
    net_{\phi_{j}}(t)&=&\sum_{m} \omega_{\phi_{j}m}y^{m}(t-1),\,y^{\phi_{j}}(t)\\
    \nonumber
    &=&f_{\phi_{j}}(net_{\phi_{j}}(t)),   
\end{eqnarray}

where $net_{\phi j}$ is the input from the network to the forget gate. The logistic
sigmoid with range $[0, 1]$ is used as squashing function $f_{\phi j}$ and weighted by the hyperbolic tangent function which has the overall task of memory correction \cite{Gers2000}. The forget gate stores all the $1$ outputs while forgetting all the $0$ outputs. Finally, LSTM can be written as \cite{Yao2015}

\begin{eqnarray}
i_t &=& \sigma(W_{xi} x_t + W_{hi} h_{t-1} + W_{ci} c_{t-1}),\\
    f_t &=& \sigma(W_{xf} x_t + W_{hf} h_{t-1} + W_{cf} c_{t-1}),\\
     o_t &=& \sigma(W_{xo} x_t + W_{ho} h_{t-1} + W_{co} c_{t-1}),\\
     h_t &=& o_t \odot \tanh(c_t).
\end{eqnarray}

Here, $i_t$
, $f_t$, and $o_t$ are input gate, forget gate and output gate of LSTM, $h_t$ represents LSTM output, $\sigma$ is LSTM logistic function, $\odot$ denotes element-wise product, $W$ is the weight metric components, $x$ is the input data in time $t$, and $c$ is LSTM memory cells.  

In our application of LSTM,
the forget gate and input gate share the same parameters, but are computed as $f_t = 1 - i_{t}$. Note that
bias terms are omitted in the above equations, but they are applied by default. A linear dependence between LSTM memory cells($c_t$) and its past($c_{t-1}$) are introduced as

\begin{equation}
    c_t = f_t \odot c_{t-1} + i_t \odot \tanh(W_{xc}x_t + W_{hc}x_{t-1}).
\end{equation}\\


\section{Hyperparameter selection}

Hyperparameter selection in neural networks is represented by optimization functions. Therefore, specifying 
hyperparameters such as the type of optimization function, learning rate, number of neurons in each 
layer, number of epochs, and validation are very important. Adam, Stochastic gradient descent(SGD), 
RMSProp, AdaDelta, and Ftrl are used as optimization functions. 

\begin{figure}
\subfloat[]{\includegraphics[width = 3in]{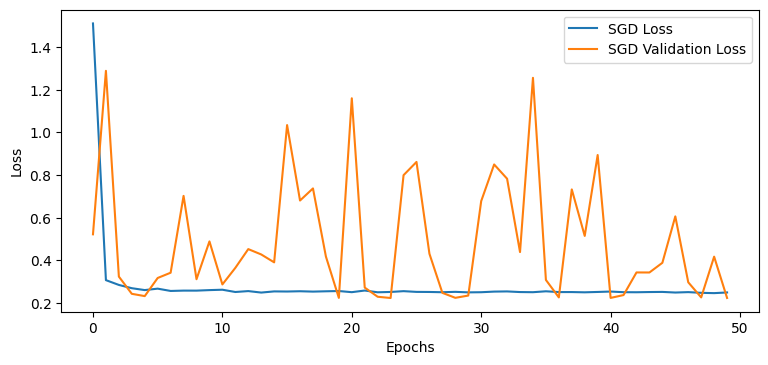}}\\
\subfloat[]{\includegraphics[width = 3in]{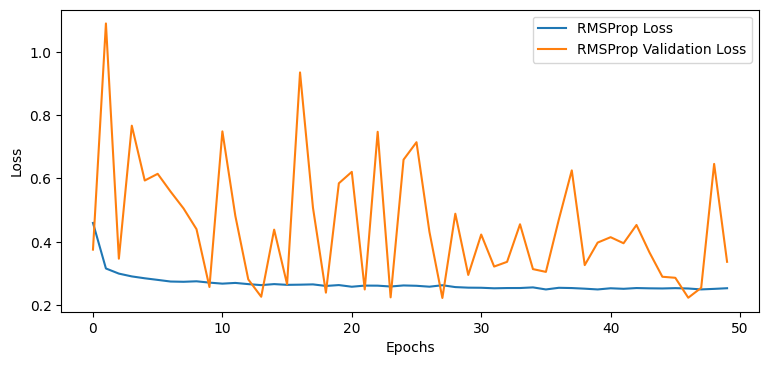}}\\
\subfloat[]{\includegraphics[width = 3in]{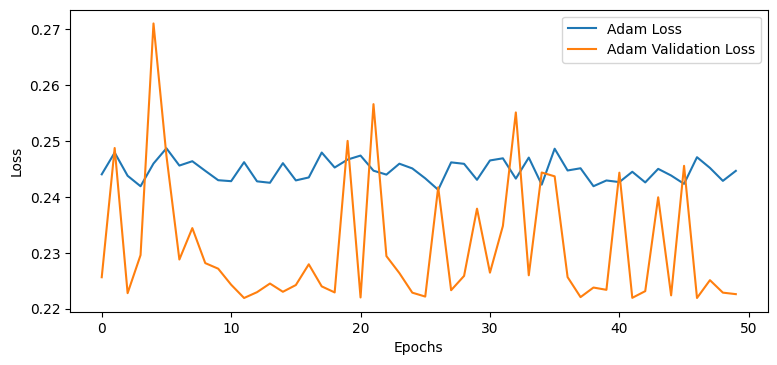}}\\
\subfloat[]{\includegraphics[width = 3in]{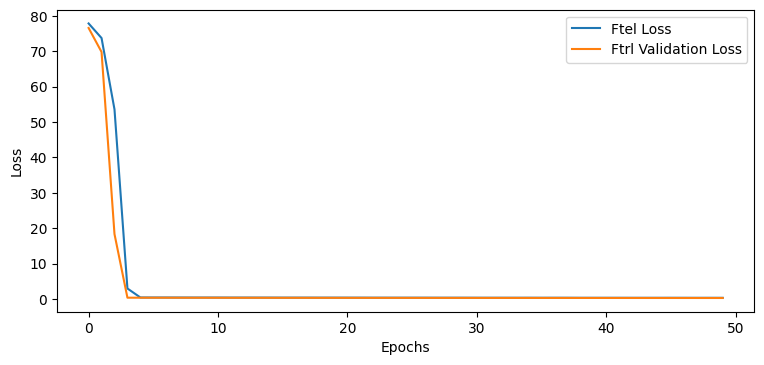}}\\
\subfloat[]{\includegraphics[width = 3in]{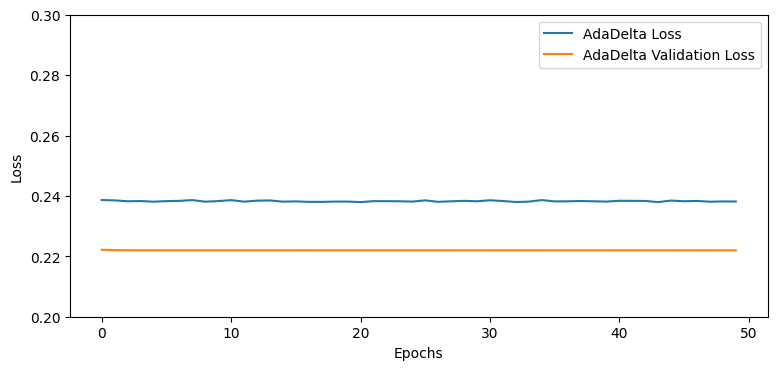}}
\caption{ \footnotesize (a) shows model evaluation for SGD optimization function. Optimization function loss is illustrated by the blue line and orange lines represent validation loss. (b) is the model evaluation using RMSProp whose optimization function loss and validation loss are shown in blue and orange. (c) illustrates the Adam model evaluation by comparing the Optimization function loss(blue line) and validation loss(orange line). The model evaluation for Ftel is shown in (d). loss of the optimization function is represented by the blue line and the validation function loss is shown by the orange line. (e) shows model evaluation for the AdaDelta optimization function. Optimization function loss is illustrated by the blue line and orange lines represent validation loss. }
\label{fig:1}
\end{figure}

We have considered about $20\%$ of the learning data 
as validation data. To determine the quality of the model, we determine the loss. The cost function that we have considered for the network is mean squared error(MSE). The number of epochs for the network learning process is equal to 
50 and the batch size is equal to 25. Results of the cost function values for each learning process 
with different optimization functions and a learning rate of $0.0005$ are shown in FIG. \ref{fig:1}.

The results related to the loss value for learning and testing data with different optimization functions are reported in the TABLE \ref{table:1}.\\

\begin{table}[h!]
\footnotesize
	\begin{center}
	\centering

\setlength{\tabcolsep}{7pt}
\renewcommand{\arraystretch}{1.5}
	\begin{tabular}{|c|c|c|}
		\hline

 		Optimization functions & Train data MSE & Test data MSE  \\ 
  \hline
			
  SGD & 0.38 & 0.39 \\
  \hline
  RMSProp & 0.37 & 0.38 \\
  \hline
  Adam & 0.23 & 0.23 \\
  \hline
  AdaDelta & 0.22 & 0.23 \\
  \hline
  Ftrl & 0.26 & 0.27 \\
  
  \hline
		\end{tabular}
		\caption{ 
		\footnotesize This table shows the result of algorithm evaluation by SGD,  RMSProp, Adam,  Ftrl and AdaDelta optimization functions.}
			\label{table:1}
	\end{center}
\end{table}


\section{Data and Network Topology}

Using QuasarNET data we predict the SMBHs mass with the help of their redshift. We use 3245 data for modelling, 2596 data for the network learning process, and 649 data for testing the network result. Data have a redshift range of 3 to 7. In the first step, data are sorted in ascending order of their redshifts. The reason is that redshift is a time series and LSTM has a recurrent architecture which creates memory through time. Then, the learning and testing data are separated in chronological order.

The network topology can be described by an LSTM layer as the dynamic layer of the network, a drop-out layer to prevent over-fitting, 3 dense layers as static layers, and the output of the network which is printed by the last dense layer. We use the hyperbolic tangent which is an active function for the LSTM layer and the first dense. Because the hyperbolic tangent is a non-linear function with a symmetric range. It is a suitable option to control sudden changes when they are in chronological order. For the second dense, we use the rectified linear unit(ReLU), to transfer the magnitude of the positive value to the next layer. For the third dense, which outputs the network as a continuous number, we use a linear function. TABLE \ref{table:2} shows the network structure based on the hyperparameters of the network.

\begin{table}[h!]
\footnotesize
	\begin{center}
 \setlength{\tabcolsep}{10pt}
 \renewcommand{\arraystretch}{1.5}
		\begin{tabular}{ | c | c |c | }
		\hline
 		Layers & Neurons & Computational Parameters  \\ 
  \hline
			
 Inputs  & - & - \\
  \hline
 LSTM  & (None,256)  & 264192
 \\
  \hline
 Dropout & (None,256) & 0 \\
  \hline
 Dense &(None,512)  & 131584\\
  \hline
 Dense  & (None,256)  & 131328\\
 \hline
 Dense  & (None,1) & 257 \\
 \hline
		\end{tabular}
  
	\begin{tabular}{c|c}
		    Total Computational Parameters & 527361 \\
		    \hline
		   Trainable
Computational Parameters &  527361 \\
\hline
Non-Trainable Computational
Parameter & 0 \\
\hline

		\end{tabular}
		\caption{ \footnotesize This table illustrates the network topology which includes each layer along with neurons and computational parameters.
		\footnotesize  }
		\label{table:2}
	\end{center}
\end{table}

One of the main challenges that always exists in ML and DL is the issue of transparency. Transparency is a dynamic issue and solving this problem is different for each task. There is no specific method to solve this problem. Many factors such as the design of an interpretable learning experience, the fundamental determination of hyperparameters by the task, the observance of the principles of feature selection, and the determination of the appropriate number of data based on characteristics can allow us to have a transparent model.

Transparency in the structure of algorithms is also noteworthy. In this paper, we investigate the transparency of the model built by the designed network. Trained data are also based on redshifts from 3 to 7. With the help of the built model, SMBHs mass at  $0<z<3$ and $7<z<10$ are then predicted.

\begin{figure}
{\includegraphics[width = 3in]{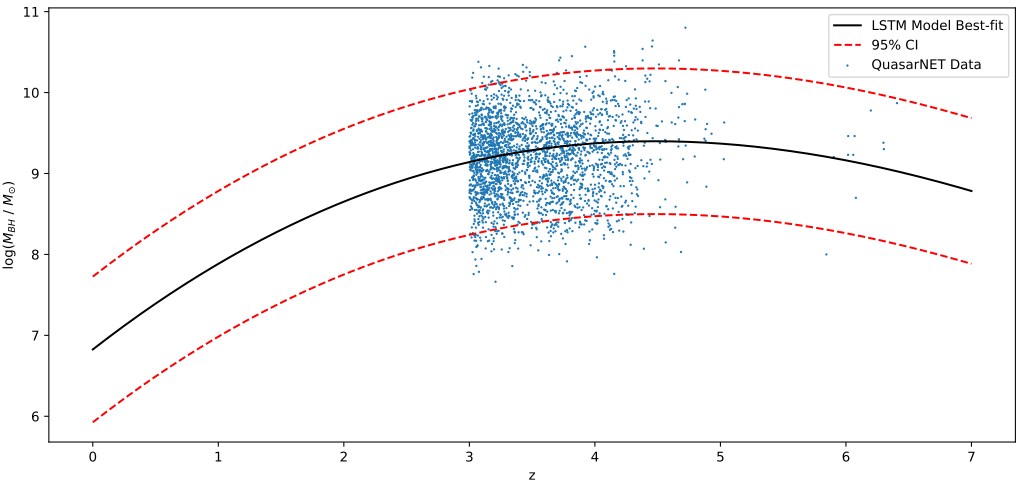}}\\
\caption{ \footnotesize Model built using flux and volume-limited samples. Corrected QuasarNET data are plotted with blue dots. The black line represents our LSTM model best-fit. In addition, red dotted lines represent our models that include 95 percent of all data.}
\label{fig:107}
\end{figure}

\begin{figure}
\subfloat[]{\includegraphics[width = 3in]{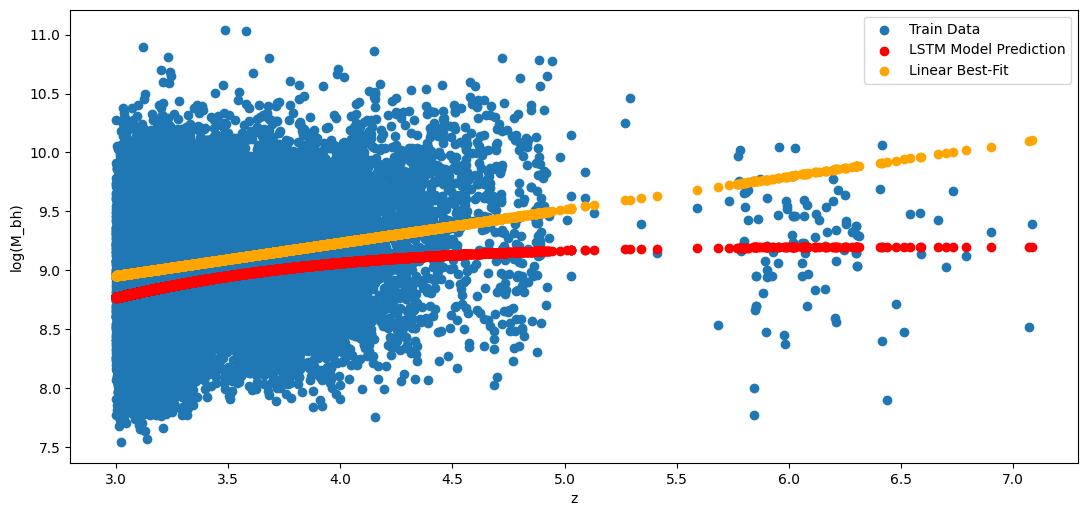}}\\
\subfloat[]{\includegraphics[width = 3in]{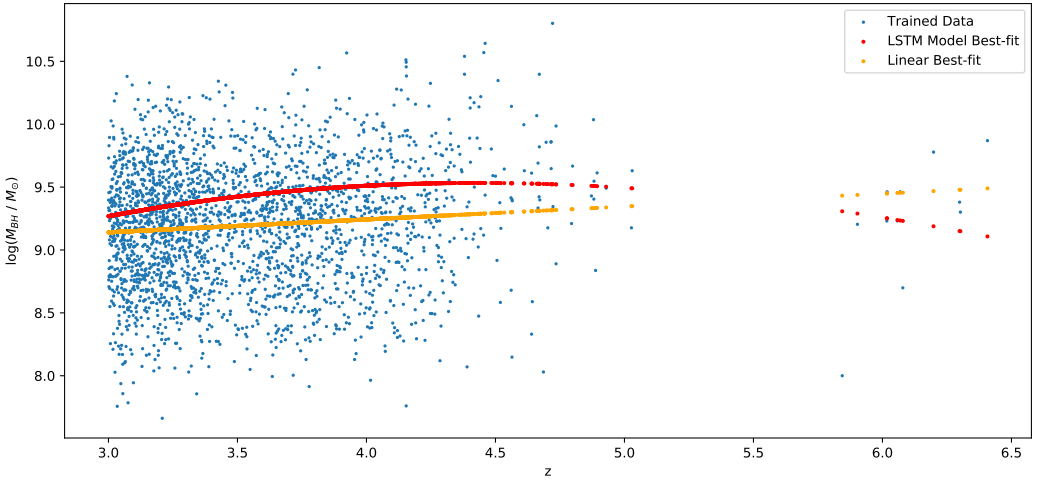}}\\
\caption{ \footnotesize (a) compares our model with the linear best-fit. Blue dots indicate train data, the LSTM model prediction is showed the colour red, and the orange line is the linear best-fit. 
(b) illustrates our model and compares it with linear best-fit based on flux and volume-limited samples. Train data is shown as blue dots, LSTM model prediction as red, and linear best-fit as an orange line.}
\label{fig:fig8}
\end{figure}

We can see the predicted changes of SMBHs mass through redshift in  FIG. \ref{fig:107} based on our built model with its 95 percent confidence level. FIG. \ref{fig:fig8} compares the linear best-fit with our LSTM model best-fit both before and after applying correction methods. It can clearly be seen that stated correcting methods change our model significantly.

\begin{figure}
\subfloat[]{\includegraphics[width = 3in]{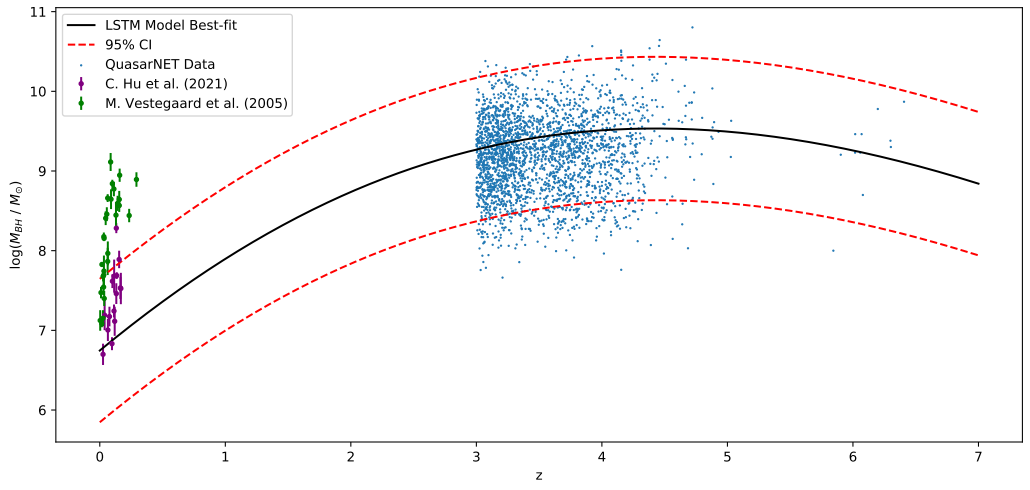}}\\
\subfloat[]{\includegraphics[width = 3in]{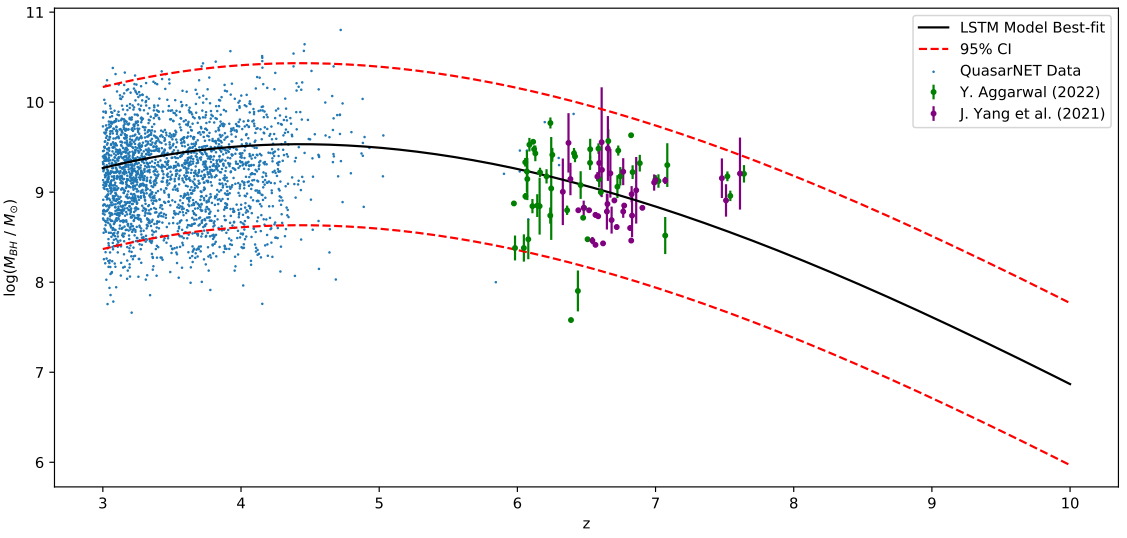}}\\
\caption{ \footnotesize (a) shows the examination of our model using multiple data sets in the redshift range of $0<z<7$. An overview of the utilized data can be found in TABLE \ref{table:3} and \ref{table:4}.
(b) is also the model examination at $3<z<10$ whose data is available in TABLE \ref{table:5} and \ref{table:6}.}
\label{fig:c}
\end{figure}

\section{Comparing with other data}

Using corrected flux and volume-limited samples of QuasarNET data, we build a DL model for quasars' central SMBH mass. 
By applying correction methods, only $\simeq 8.62\%$ of QuasarNET data is accepted to use for modelling. FIG. \ref{fig:108} shows the accepted data along with the removed data. 

Moreover, FIG. \ref{fig:107} illustrated our model whose best-fit contains 95 percent of corrected data samples. 
The model shows that SMBHs mass increases in $0<z<4.72$ and reaches its peak at $z\simeq4.72$. The mass then falls exponentially with increasing redshift at $z>4.72$. It should be noted that our model yields a different result than what is shown in other recent works like \cite{Benny2019}, where the peak is $z<4$. Nevertheless, in some studies which attempt to show quasars' central SMBHs mass evolution,  Eq.  \ref{eqn:eq115} is used that does not include any peaks(e.g. see \cite{Banados2017}).

The model is then evaluated by using different data sets which are available in multiple tables. We use the results of the long-term spectroscopic monitoring of 15 PG quasars that have relatively strong Fe II emission to generate TABLE \ref{table:3} \cite{Hu2021}. Moreover, TABLE \ref{table:4} shows relatively nearby quasars with redshifts obtained from the NED and central SMBHs mass determined through multi-epoch spectrophotometry and RM \cite{Vestergaard2005}. A list of 69 high-redshift quasars is also available in TABLE \ref{table:5} and TABLE \ref{table:6}. For each quasar, the most accurate estimation of its central SMBH mass using Mg II emission lines along with its uncertainty is shown \cite{Aggarwal2022, Yang2021}. While the model matches observational data quite well at $3<z<10$, there is a minor deviation at lower redshifts, i.e. $0<z<3$. The comparison between our model predictions and the observational data for both low-redshift and high-redshift quasars can be seen in FIG. \ref{fig:c}.

In addition to gas being sucked into SMBHs, there is an alternative process that turns them into stars. There has been a comparison of SMBH accretion rate and SFR on a galactic scale in several observational studies \cite{Netzer2007, Wild2007, Wild2010, Rosario2012}.

In our next work, we will address the SFR and its effects on the model. Thus, it is possible to fix the minor deviation between the model and observations. Further, there are more data available for lower redshift quasars, compared to higher ones, whose reasons should be studied and may have an impact on the final results of our model.\\


\section{Conclusions}
The question of how the SMBHs that have been observed in the universe came into being is one of the biggest questions in cosmology. In recent years, it has been established that stellar BHs cannot accrete mass, resulting in such BHs. If we want to consider these BHs as stellar BHs that have reached such incredible mass due to accretion, the age of the universe should have been much longer than it is. On the other hand, it is impossible for a star to form a SMBH as a result of its collapse. In addition, there is another idea that states that these BHs are actually primordial BHs. Although this idea is very controversial, it has not been rejected yet. There are even hopes to prove such a thing.

One of the most interesting surveys available for quasars is the SDSS. In this paper, we have used SDSS-DR16Q. In particular, we have taken advantage of the QuasarNET research platform. QuasarNET specifically has focused on the study of SMBHs. Although 37648 data in redshifts between 3 and 7 have been reported in it, these data need accurate corrections to be used. These corrections are flux and volume-limited, which makes the right conditions to work on SMBHs over time for training the machine. After applying these corrections, 3246 data remained and 34403 data were removed. In FIG. \ref{fig:108} we have plotted accepted and removed data after correcting them.

Considering the remaining 3246 data of the mass of BHs in the center of quasars at redshifts between 3 and 7, we have modeled them over time with the help of the LSTM RNN. We have elaborated details of our used DL approach in several sections. The model we have presented with the help of QuasarNET data tries to predict the mass of the central massive BHs of quasars at redshifts between 0 and 10.

Firstly, in FIG. \ref{fig:fig8}, we have compared our prediction with the linear best-fit of QuasarNET data before and after correcting data. Then, we illustrated the best-fit and a band that 95 percent of the QuasarNET data is within 2 standard deviations of the mean for our model in redshifts 0 to 10.

Eventually, we should have compared our model with other observational data at redshifts between 0 and 3 and also 7 and 10. This will enable us to see whether our model works or not. We have used four data sets for this comparison. Two of them are related to redshifts 0 to 3 and the other two are related to redshifts 7 to 10. FIG. \ref{fig:c} demonstrates two redshift ranges, 0 to 7 and 3 to 10. As it is evident, at redshifts higher than 7, our model has a very good description of the data and can make a reliable prediction, but at redshifts below 3, it seems that there is a slight deviation.

This deviation can be due to not considering other parameters describing quasars. We have only used the estimation of the mass of the central SMBHs of quasars and their redshift in QuasarNET data. However, data such as the Eddington ratio and bolometric luminosity are also available and can be used for subsequent modeling.

Another thing that can improve the model is to consider star formation with the help of other observational data sets. Accurately obtaining the time of star formation causes the redshift of the peak of the model we obtained to change to lower redshifts. This issue makes our model predict more massive central SMBHs at redshifts below 3, and as a result, it fits better with other data.

Finally, we must state that this effort to model SMBHs at high redshifts will help us to find out when and how they have been formed and their role in the formation of the structures. Furthermore, if the process of their growth through the accretion and merger of primordial BHs is also studied in future works, it will probably yield interesting results. Because by going back through time, the initial masses of these central SMBHs can be examined.

\section*{Acknowledgement}
Authors thank Shant Baghram for the great discussions that helped us to model and correct the QuasarNET data and Rahim Moradi for helpful discussion.

\section*{Data availability}
The catalogue underlying this paper is available in the Sloan
Digital Sky Survey Quasar catalogue: 16th data release
(DR16Q) at \url{https://www.sdss.org/dr16/algorithms/qsocatalog/} \cite{Lyke2020}.

The data that support the findings of this study are openly available at \url{https://www.kaggle.com/datasets/quasarnet/quasarnet}, reference number \cite{QuasarNet}.


\newpage
\begin{thebibliography}{99}

\bibitem{inayoshi2019assembly}
Inayoshi, Kohei, Eli Visbal, and Zoltán Haiman. "The assembly of the first massive black holes." arXiv preprint arXiv:1911.05791 (2019).

\bibitem{blandford1982reverberation}
Blandford, R. D., and C. F. McKee. "Reverberation mapping of the emission line regions of Seyfert galaxies and quasars." The Astrophysical Journal 255 (1982): 419-439.

\bibitem{du2019radius}
Du, Pu, and Jian-Min Wang. "The radius–luminosity relationship depends on optical spectra in active galactic nuclei." The Astrophysical Journal 886.1 (2019): 42.

\bibitem{antonucci1993unified}
Antonucci, Robert. "Unified models for active galactic nuclei and quasars." Annual review of astronomy and astrophysics 31 (1993): 473-521.

\bibitem{shields1978thermal}
Shields, G. A. "Thermal continuum from accretion disks in quasars." Nature 272.5655 (1978): 706-708.

\bibitem{svensson1994black}
Svensson, Roland, and Andrzej A. Zdziarski. "Black hole accretion disks with coronae." The Astrophysical Journal 436 (1994): 599-606.

\bibitem{arnaud1985exosat}
Arnaud, K. A., et al. "EXOSAT observations of a strong soft X-ray excess in MKN 841." Monthly Notices of the Royal Astronomical Society 217.1 (1985): 105-113.

\bibitem{yang2020poniua}
Yang, Jinyi, et al. "Pōniuā ‘ena: A Luminous z= 7.5 Quasar Hosting a 1.5 Billion Solar Mass Black Hole." The Astrophysical Journal Letters 897.1 (2020): L14.

\bibitem{wise2019formation}
Wise, John H., et al. "Formation of massive black holes in rapidly growing pre-galactic gas clouds." Nature 566.7742 (2019): 85-88.

\bibitem{kroupa2020very}
Kroupa, Pavel, et al. "Very high redshift quasars and the rapid emergence of supermassive black holes." Monthly Notices of the Royal Astronomical Society 498.4 (2020): 5652-5683.

\bibitem{bernal2018signatures}
Bernal, José Luis, et al. "Signatures of primordial black holes as seeds of supermassive black holes." Journal of Cosmology and Astroparticle Physics 2018.05 (2018): 017.

\bibitem{volonteri2015case}
Volonteri, Marta, Joseph Silk, and Guillaume Dubus. "The case for supercritical accretion onto massive black holes at high redshift." The Astrophysical Journal 804.2 (2015): 148.


\bibitem{Lyke2020}
Lyke, Brad W., et al. "The Sloan Digital Sky Survey Quasar Catalog: Sixteenth Data Release." The Astrophysical Journal Supplement Series 250.1 (2020): 8.


\bibitem{Busca2018}
Busca, Nicolas, and Christophe Balland. "QuasarNET: Human-level spectral classification and redshifting with Deep Neural Networks." arXiv preprint arXiv:1808.09955 (2018).

\bibitem{ball2010data}
Ball, Nicholas M., and Robert J. Brunner. "Data mining and machine learning in astronomy." International Journal of Modern Physics D 19.07 (2010): 1049-1106.

\bibitem{baron2019machine}
Baron, Dalya. "Machine learning in astronomy: A practical overview." arXiv preprint arXiv:1904.07248 (2019).

\bibitem{allen2019deep}
Allen, Gabrielle, et al. "Deep learning for multi-messenger astrophysics: A gateway for discovery in the big data era." arXiv preprint arXiv:1902.00522 (2019).

\bibitem{meher2021deep}
Meher, Saroj K., and Ganapati Panda. "Deep learning in astronomy: a tutorial perspective." The European Physical Journal Special Topics 230.10 (2021): 2285-2317.

\bibitem{nakoneczny2021photometric}
Nakoneczny, S. J., et al. "Photometric selection and redshifts for quasars in the Kilo-Degree Survey Data Release 4." Astronomy and Astrophysics 649 (2021): A81.

\bibitem{wenzl2021random}
Wenzl, Lukas, et al. "Random forests as a viable method to select and discover high-redshift quasars." The Astronomical Journal 162.2 (2021): 72.

\bibitem{burhanudin2021light}
Burhanudin, U. F., et al. "Light-curve classification with recurrent neural networks for GOTO: dealing with imbalanced data." Monthly Notices of the Royal Astronomical Society 505.3 (2021): 4345-4361.

\bibitem{vardoulaki2021fr}
Vardoulaki, E., et al. "FR-type radio sources at 3 GHz VLA-COSMOS: Relation to physical properties and large-scale environment." Astronomy and Astrophysics 648 (2021): A102.

\bibitem{wang2022j}
Wang, Cunshi, et al. "J-PLUS: Support vector machine applied to STAR-GALAXY-QSO classification." Astronomy and Astrophysics 659 (2022): A144.

\bibitem{xiao2020efficient}
Xiao, H. B., et al. "Efficient Fermi source identification with machine learning methods." Astronomy and Computing 32 (2020): 100387.

\bibitem{parkinson2016classification}
Parkinson, PM Saz, et al. "Classification and ranking of Fermi LAT gamma-ray sources from the 3FGL catalog using machine learning techniques." The Astrophysical Journal 820.1 (2016): 8.

\bibitem{li2021ai}
Li, Yin, et al. "AI-assisted superresolution cosmological simulations." Proceedings of the National Academy of Sciences 118.19 (2021): e2022038118.


\bibitem{dainotti2021predicting}
Narendra, Aditya, et al. "Predicting the redshift of gamma-ray loud quasars using Supervised Machine Learning: Part 2." arXiv preprint arXiv:2201.05374 (2022).

\bibitem{York2000}
York, Donald G., et al. "The sloan digital sky survey: Technical summary." The Astronomical Journal 120.3 (2000): 1579.
\bibitem{Rastegarnia2022}
Rastegarnia, F., et al. "Deep learning in searching the spectroscopic redshift of quasars." Monthly Notices of the Royal Astronomical Society 511.3 (2022): 4490-4499.


\bibitem{Chen2005}
Xie, G.-Z., Chen, L.-E., Xie, Z.-H., Ma, L., and Zhou, S.-B. (2005). Agn black hole masses
and methods to estimate the mass. Publications of the Astronomical Society of Japan,
57(1):183–186.

\bibitem{Vestergaard2002}
Vestergaard, Marianne. "Determining central black hole masses in distant active galaxies." The Astrophysical Journal 571.2 (2002): 733.

\bibitem{Wandel1999}
Wandel, A., B. M. Peterson, and M. A. Malkan. "Central masses and broad-line region sizes of active galactic nuclei. I. Comparing the photoionization and reverberation techniques." The Astrophysical Journal 526.2 (1999): 579.

\bibitem{Rodriguez1997}
Rodriguez-Pascual, P. M., et al. "Steps toward determination of the size and structure of the broad-Line region in active galactic nuclei. IX. Ultraviolet observations of fairall 9." The Astrophysical Journal Supplement Series 110.1 (1997): 9.

\bibitem{Bentz2009}
Bentz, Misty C., et al. "The Lick AGN monitoring project: Broad-line region radii and black hole masses from reverberation mapping of H$\beta$." The Astrophysical Journal 705.1 (2009): 199.

\bibitem{Netzer2013}
Netzer, Hagai. The physics and evolution of active galactic nuclei. Cambridge university press, 2013.


\bibitem{McLure2004}
McLure, Ross J., and James S. Dunlop. "The cosmological evolution of quasar black hole masses." Monthly Notices of the Royal Astronomical Society 352.4 (2004): 1390-1404.

\bibitem{McLure2002}
McLure, Ross J., and Matt J. Jarvis. "Measuring the black hole masses of high-redshift quasars." Monthly Notices of the Royal Astronomical Society 337.1 (2002): 109-116.



\bibitem{Wills1986}
Wills, B. J, and I. W. A. Browne. "Relativistic beaming and quasar emission lines." The Astrophysical Journal 302 (1986): 56-63.

\bibitem{Koss2017}
Koss, Michael, et al. "BAT AGN spectroscopic survey. I. Spectral measurements, derived quantities, and AGN demographics." The Astrophysical Journal 850.1 (2017): 74.

\bibitem{Schutte2019}
Schutte, Zachary, Amy E. Reines, and Jenny E. Greene. "The black hole–bulge mass relation including dwarf galaxies hosting active galactic nuclei." The Astrophysical Journal 887.2 (2019): 245.

\bibitem{Willott2010}
Willott, Chris J., et al. "Eddington-limited accretion and the black hole mass function at redshift 6." The Astronomical Journal 140.2 (2010): 546.


\bibitem{Fanidakis2012}
Fanidakis, N., et al. "The evolution of active galactic nuclei across cosmic time: what is downsizing?." Monthly Notices of the Royal Astronomical Society 419.4 (2012): 2797-2820.

\bibitem{Hopkins2007}
Hopkins, Philip F., Gordon T. Richards, and Lars Hernquist. "An observational determination of the bolometric quasar luminosity function." The Astrophysical Journal 654.2 (2007): 731.

\bibitem{Benny2019}
Trakhtenbrot, Benny. "What do observations tell us about the highest-redshift supermassive black holes?." Proceedings of the International Astronomical Union 15.S356 (2019): 261-275.

\bibitem{Benny2017}
Trakhtenbrot, Benny, Marta Volonteri, and Priyamvada Natarajan. "On the accretion rates and radiative efficiencies of the highest-redshift quasars." The Astrophysical Journal Letters 836.1 (2017): L1.

\bibitem{Natarajan2021}
Natarajan, Priyamvada, et al. "QuasarNet: A new research platform for the data-driven investigation of black holes." arXiv preprint arXiv:2103.13932 (2021).

\bibitem{Salvato2019}
Salvato, Mara, Olivier Ilbert, and Ben Hoyle. "The many flavours of photometric redshifts." Nature Astronomy 3.3 (2019): 212-222.

\bibitem{Tago2010}
Tago, E., et al. "Groups of galaxies in the SDSS Data Release 7-Flux-and volume-limited samples." Astronomy and Astrophysics 514 (2010): A102.





lags." The Astrophysical Journal 856.1 (2018): 6.


\bibitem{Staudemeyer2019}
Staudemeyer, Ralf C., and Eric Rothstein Morris. "Understanding LSTM--a tutorial into long short-term memory recurrent neural networks." arXiv preprint arXiv:1909.09586 (2019).

\bibitem{Williams1990}
Williams, Ronald J., and David Zipser. Gradient-based learning algorithms for recurrent connectionist networks. Boston, MA: College of Computer Science, Northeastern University, 1990.

\bibitem{Werbos1990}
Werbos, Paul J. "Backpropagation through time: what it does and how to do it." Proceedings of the IEEE 78.10 (1990): 1550-1560.

\bibitem{Bengio1994}
Bengio, Yoshua, Patrice Simard, and Paolo Frasconi. "Learning long-term dependencies with gradient descent is difficult." IEEE transactions on neural networks 5.2 (1994): 157-166.

\bibitem{Hochreiter1996}
Hochreiter, Sepp, and Jürgen Schmidhuber. "LSTM can solve hard long time lag problems." Advances in neural information processing systems 9 (1996).

\bibitem{Gers2000}
Gers, Felix A., Jürgen Schmidhuber, and Fred Cummins. "Learning to forget: Continual prediction with LSTM." Neural computation 12.10 (2000): 2451-2471.

\bibitem{Gers2002}
Gers, Felix A., Nicol N. Schraudolph, and Jürgen Schmidhuber. "Learning precise timing with LSTM recurrent networks." Journal of machine learning research 3.Aug (2002): 115-143.

\bibitem{Hochreiter1997}
Hochreiter, Sepp, and Jürgen Schmidhuber. "Long short-term memory." Neural computation 9.8 (1997): 1735-1780.

\bibitem{Graves2012}
Graves, Alex. "Long short-term memory." Supervised sequence labelling with recurrent neural networks (2012): 37-45.

\bibitem{Yao2015}
Yao, Kaisheng, et al. "Depth-gated LSTM." arXiv preprint arXiv:1508.03790 (2015).


\bibitem{Banados2017}
Banados, Eduardo, et al. "An 800 million solar mass black hole in a significantly neutral universe at redshift 7.5." arXiv preprint arXiv:1712.01860 (2017).


\bibitem{Hu2021}
Hu, Chen, et al. "Supermassive Black Holes with High Accretion Rates in Active Galactic Nuclei. XII. Reverberation Mapping Results for 15 PG Quasars from a Long-duration High-cadence Campaign." The Astrophysical Journal Supplement Series 253.1 (2021): 20.

\bibitem{Vestergaard2005}
Vestergaard, P., L. Rejnmark, and L. Mosekilde. "Relative fracture risk in patients with diabetes mellitus, and the impact of insulin and oral antidiabetic medication on relative fracture risk." Diabetologia 48.7 (2005): 1292-1299.

\bibitem{Aggarwal2022}
Aggarwal, Yash. "New insights into the origins and growth of seeds of supermassive black holes."

\bibitem{Yang2021}
Yang, Jinyi, et al. "Probing Early Supermassive Black Hole Growth and Quasar Evolution with Near-infrared Spectroscopy of 37 Reionization-era Quasars at 6.3 z 7.64." The Astrophysical Journal 923.2 (2021): 262.


\bibitem{Netzer2007}
Netzer, Hagai, et al. "Spitzer quasar and ULIRG evolution study (QUEST). II. The spectral energy distributions of palomar-green quasars." The Astrophysical Journal 666.2 (2007): 806.

\bibitem{Wild2007}
Wild, Vivienne, et al. "Bursty stellar populations and obscured active galactic nuclei in galaxy bulges." Monthly Notices of the Royal Astronomical Society 381.2 (2007): 543-572.

\bibitem{Wild2010}
Wild, Vivienne, Timothy Heckman, and Stéphane Charlot. "Timing the starburst–AGN connection." Monthly Notices of the Royal Astronomical Society 405.2 (2010): 933-947.

\bibitem{Rosario2012}
Rosario, D. J., et al. "The mean star formation rate of X-ray selected active galaxies and its evolution from z~ 2.5: results from PEP-Herschel." Astronomy and Astrophysics 545 (2012): A45.

\bibitem{QuasarNet}
QuasarNet (2022). QuasarNet [Dataset]. https://www.kaggle.com/datasets/quasarnet/quasarnet


\bibitem{Wang2021}
Wang, Feige, et al. "A luminous quasar at redshift 7.642." The Astrophysical Journal Letters 907.1 (2021): L1.

a redshift of 7.5, Nature 553 473 (2017)

\bibitem{Daniel2011}
Mortlock, Daniel J., et al. "A luminous quasar at a redshift of z= 7.085." Nature 474.7353 (2011): 616-619.

\bibitem{Matsuoka2019}
Matsuoka, Yoshiki, et al. "Discovery of the First Low-luminosity Quasar at z> 7." The Astrophysical Journal Letters 872.1 (2019): L2.

\bibitem{Wang2018}
Wang, Feige, et al. "The discovery of a luminous broad absorption line quasar at a redshift of 7.02." The Astrophysical Journal Letters 869.1 (2018): L9.

\bibitem{Wang2020}
Wang, Feige, et al. "A Significantly Neutral Intergalactic Medium Around the Luminous z 7 Quasar J0252–0503." The Astrophysical Journal 896.1 (2020): 23.

\bibitem{Venemans2013}
B.P. Venemans et al., Discovery of Three z 6.5 Quasars in the VISTA Kilo-Degree Infrared 
Galaxy (VIKING) Survey, ApJ 779 24 (2013)

\bibitem{Reed2019}
Reed, Sophie L., et al. "Three new VHS–DES quasars at 6.7< z< 6.9 and emission line properties at z> 6.5." Monthly Notices of the Royal Astronomical Society 487.2 (2019): 1874-1885.


\bibitem{Matsuoka2016}
Matsuoka, Yoshiki, et al. "SUBARU HIGH-z EXPLORATION OF LOW-LUMINOSITY QUASARS (SHELLQs). I. DISCOVERY OF 15 QUASARS AND BRIGHT GALAXIES." The Astrophysical Journal 828.1 (2016): 26.

\bibitem{Mazzucchelli2017}
Mazzucchelli, C., et al. "Physical properties of 15 quasars ." The Astrophysical Journal 849.2 (2017): 91.


\bibitem{Eilers2020}
Eilers, Anna-Christina, et al. "Detecting and characterizing young quasars. I. Systemic redshifts and proximity zone measurements." The Astrophysical Journal 900.1 (2020): 37.

\bibitem{Onoue2020}
Onoue, Masafusa, et al. "No Redshift Evolution in the Broad-line-region Metallicity up to z= 7.54: Deep Near-infrared Spectroscopy of ULAS J1342+ 0928." The Astrophysical Journal 898.2 (2020): 105.

\bibitem{Mortlock2009}
Mortlock, D. J., et al. "Discovery of a redshift 6.13 quasar in the UKIRT infrared deep sky survey." Astronomy and Astrophysics 505.1 (2009): 97-104.

\end{thebibliography}


\begin{table*}[p]
\footnotesize
	\begin{center}
\begin{tabularx}{1.0\textwidth} { 
  | >{\centering\arraybackslash}X 
  | >{\centering\arraybackslash}X 
  | >{\centering\arraybackslash}X|}
 \hline
 Object & Redshift  & $M_{BH}(\times 10^7 M_{\odot})$  \\
 \hline
 PG 0003+199  & 0.0259  & $0.50^{+0.18}_{-0.18}$\\
PG  0804+761  & 0.1005  & $4.14^{+0.91}_{-0.98}$\\
PG   0838+770 & 0.1316  & $2.89^{+1.01}_{-1.13}$\\
PG  1115+407 & 0.1542  & $7.76^{+2.23}_{-1.95}$\\
PG   1322+659 & 0.1678  & $3.35^{+1.92}_{-0.71}$\\
PG 1402+261  & 0.1643  & $3.41^{+1.28}_{-1.51}$\\
PG   1404+226  & 0.0972  & $0.68^{+0.14}_{-0.23}$\\
PG  1415+451 & 0.1132  & $1.75^{+0.36}_{-0.32}$\\
PG   1440+356  & 0.0770  & $1.49^{+0.49}_{-0.55}$\\
PG   1448+273 & 0.0646  & $1.01^{+0.38}_{-0.23}$\\
PG    1519+226  &0.1351  & $4.87^{+0.49}_{-0.86}$\\
PG    1535+547  &0.0385  & $1.55^{+0.84}_{-0.82}$\\
PG     1552+085 & 0.1187  & $1.30^{+0.68}_{-0.65}$\\
PG    1617+175 & 0.1144  & $4.79^{+2.94}_{-2.83}$\\
PG    1626+554  &0.1316  & $19.17^{+2.98}_{-2.73}$\\
\hline
\end{tabularx}
		\caption{
		This table contains 15 low redshift quasars at $z<1$ with their central SMBH mass reported in \cite{Hu2021}.}
			\label{table:3}
	\end{center}
\end{table*}

\begin{table*}[p]
\footnotesize
	\begin{center}
\begin{tabularx}{1.0\textwidth} { 
  | >{\centering\arraybackslash}X 
  | >{\centering\arraybackslash}X 
  | >{\centering\arraybackslash}X|}
 \hline
 Object & Redshift  & $\log(M/M_\odot)\,(H\beta, rms)$\\
 \hline
 Mrk 335   & 0.02578 & $7.152^{+0.101}_{-0.131}$\\
PG 0026+129  &0.14200  & $8.594^{+0.095}_{-0.122}$ \\
PG 0052+251 & 0.15500 & $8.567^{+0.081}_{-0.100}$  \\
Fairall 9 & 0.04702 & $8.407^{+0.086}_{-0.108}$\\
Mrk 590 & 0.02638  & $7.677^{+0.063}_{-0.074}$\\
3C 120  & 0.03301  & $7.744^{+0.195}_{-0.226}$\\
Ark 120 & 0.03230  & $8.176^{+0.052}_{-0.059}$\\
PG 0804+761 & 0.10000  & $8.841^{+0.049}_{-0.055}$\\
PG 0844+349  & 0.06400  & $7.966^{+0.150}_{-0.231}$\\
Mrk 110 & 0.03529  & $7.400^{+0.094}_{-0.121}$\\
PG 0953+414  &0.23410  & $8.441^{+0.084}_{-0.104}$\\
NGC 3783  &0.00973 & $7.474^{+0.072}_{-0.087}$\\
NGC 4151 &0.00332 & $7.124^{+0.129}_{-0.184}$\\
PG 1226+023 & 0.15830 & $8.947^{+0.083}_{-0.103}$\\
PG 1229+204  &0.06301 & $7.865^{+0.171}_{-0.285}$\\
PG 1307+085  &0.15500 & $8.643^{+0.107}_{-0.142}$\\
Mrk 279   &0.03045 & $7.543^{+0.102}_{-0.133}$\\
PG 1411+442  &0.08960 & $8.646^{+0.124}_{-0.174}$\\
NGC 5548  &0.01717  & $7.827^{+0.017}_{-0.017}$\\
PG 1426+015  &0.08647  & $9.113^{+0.113}_{-0.153}$\\
Mrk 817  &0.03145  & $7.694^{+0.063}_{-0.074}$\\
PG 1613+658 &0.12900  & $8.446^{+0.165}_{-0.270}$\\
PG 1617+175 &0.11240  & $8.774^{+0.019}_{-0.115}$\\
PG 1700+518 &0.29200  & $8.893^{+0.091}_{-0.103}$\\
3C 390.3 &0.05610 & $8.458^{+0.087}_{-0.110}$\\
Mrk 509 &0.03440  & $8.115^{+0.035}_{-0.038}$\\
PG 2130+099 &0.06298  & $8.660^{+0.049}_{-0.056}$\\
NGC 7469 &0.01632 & $7.086^{+0.047}_{-0.053}$\\
\hline
\end{tabularx}
		\caption{ 
		This table contains 28 low redshift quasars at $z<1$ with their central SMBH mass from \cite{Vestergaard2005}.}
			\label{table:4}
	\end{center}
\end{table*}

\begin{table*}[p]
\footnotesize
	\begin{center}
\begin{tabularx}{1\textwidth} { 
  | >{\centering\arraybackslash}X 
  | >{\centering\arraybackslash}X 
  | >{\centering\arraybackslash}X
  | >{\centering\arraybackslash}X|}
 \hline

 Object & Redshift & $M_{BH} (\times 10^9 M_\odot)$& Refence\\
 \hline
 J0313-1806   & 7.64 & $0.16^{+0.4}_{-0.4}$ &  \cite{Wang2021} \\
 ULAS J1342+0928   &7.541  & $0.91^{+0.13}_{-0.14}$ &  \cite{Banados2017}\\
 J100758.264+211529.207 &7.52& $1.5^{+0.2}_{-0.2}$ &  \cite{yang2020poniua}\\
 ULAS J1120+0641 & 7.085& $2.0^{+1.5}_{-0.7}$ &  \cite{Daniel2011}\\
 J124353.93+010038.5 & 7.07& $0.33^{+0.2}_{-0.2}$ &  \cite{Matsuoka2019}\\
 J0038-1527  & 7.021 & $1.33^{+0.25}_{-0.25}$ & \cite{Wang2018}\\
 DES J025216.64–050331.8 & $7$ & $1.39^{+0.16}_{-0.16}$ &  \cite{Wang2020}\\
 ULAS J2348-3054 & 6.886 & $2.1^{+0.5}_{-0.5}$ &  \cite{Venemans2013}\\
 VDES J0020-3653  & 6.834 & $1.67^{+0.32}_{-0.32}$ &  \cite{Reed2019}\\
 PSO J172.3556+18.7734  & 6.823& $3.7^{+1.3}_{-1.0}$ &  \cite{Venemans2013}\\
 ULAS J0109-3047 &6.745& $1.0^{+0.1}_{-0.1}$ & \cite{Venemans2013}\\
 HSC J1205-0000 &6.73& $1.15^{+0.39}_{-0.39}$ &  \cite{Reed2019}\\
 VDES J0244-5008 &6.724& $3.7^{+1.3}_{-1.0}$ &  \cite{Venemans2013}\\
 PSO J338.2298 & 6.658& $3.7^{+1.3}_{-1.0}$ &  \cite{Venemans2013}\\
 ULAS J0305-3150  &6.604 & $1.0^{+0.1}_{-0.1}$&  \cite{Venemans2013}\\
 PSO J323.1382  & 6.592 & $1.39^{+0.32}_{-0.51}$ &  \cite{Mazzucchelli2017}\\
 PSO J231.6575 &6.587& $3.05^{+0.44}_{-2.24}$ &  \cite{Mazzucchelli2017}\\
 PSO J036.5078& 6.527 & $3^{+0.92}_{-0.77}$ &  \cite{Mazzucchelli2017}\\
 $VDES J0224-4711$& 6.526& $2.12^{+0.42}_{-0.42}$ &  \cite{Reed2019}\\
 $PSO J167.6415$&6.508 & $0.3^{+0.008}_{-0.012}$ &  \cite{Venemans2013}\\
 $PSO J261+19$&6.483 & $0.67^{+0.21}_{-0.21}$ &  \cite{Eilers2020}\\
 $PSO J247.2970$& 6.476  & $5.2^{+0.22}_{-0.25}$ &  \cite{Mazzucchelli2017}\\
 $PSO J011+09 $& 6.458  & $1.2^{+0.51}_{-0.51}$ &  \cite{Eilers2020}\\
 $CFHQS J0210-0456$& 6.438  & $0.08^{+0.055}_{-0.04}$ &  \cite{Willott2010}\\
 $CFHQS J2329-0301$ &6.417& $2.5^{+0.4}_{-0.4}$ &  \cite{Willott2010}\\
 $HSC J0859 +0022$ & 6.388& $0.038^{+0.001}_{-0.0018}$ &  \cite{Matsuoka2016}\\
 $HSC J2239 +0207$ &6.245&$1.1^{+3}_{-2}$ &  \cite{Mazzucchelli2017}\\
 $VDES J0330–4025$ &6.239&$5.87^{+0.89}_{-0.89}$ &  \cite{Eilers2020}\\
 $VDES J0323–4701$  &6.238&$0.55^{+0.126}_{-0.126}$ &  \cite{Eilers2020}\\
 $PSO J359–06$ &6.164&$1.66^{+0.21}_{-0.21}$ &  \cite{Eilers2020}\\
 $CFHQS J0221-0802$ &6.161&$0.7^{+0.75}_{-0.47}$ &  \cite{Willott2010}\\
 $HSC J1208-0200$ &6.144&$0.71^{+0.24}_{-0.52}$ & \cite{Onoue2020}\\
 $ULAS J1319+0950$ &6.13& $2.7^{+0.6}_{-0.6}$ & \cite{Mortlock2009}\\
 $CFHQS J1509-1749$ &6.121& $3.0^{+0.3}_{-0.3}$ &  \cite{Willott2010}\\
$PSO J239–07$  &6.114 &$3.63^{+0.2}_{-0.2}$ &  \cite{Eilers2020}\\
 $HSC J2216-0016$ &6.109& $0.7^{+0.14}_{-0.23}$ &  \cite{Onoue2020}\\
 $CFHQS J2100-1715$  &6.087&$3.37^{+0.64}_{-0.64}$ &  \cite{Willott2010}\\
 $PSO J158–14$ &6.057&$2.15^{+0.25}_{-0.25}$ & \cite{Eilers2020}\\
 $CFHQS J1641+3755$ &6.047&$0.24^{+0.1}_{-0.8}$ &  \cite{Willott2010}\\
 $CFHQS J0055+0146$ & 5.983&$0.24^{+0.9}_{-0.7}$ &  \cite{Willott2010}\\
 $PSO J056–16$ & 5.975 & $0.75^{+0.007}_{-0.007}$&  \cite{Eilers2020}\\
\hline
    \end{tabularx}
		\caption{ 
		This table contains 41 high-redshift quasars at $z>5$ with their central SMBH mass from different references which are identified in the fourth column.}
			\label{table:5}
	\end{center}
 \end{table*}

\begin{table*}[p]
\footnotesize
	\begin{center}
\begin{tabularx}{1.0\textwidth} { 
  | >{\centering\arraybackslash}X 
  | >{\centering\arraybackslash}X 
  | >{\centering\arraybackslash}X|}
 \hline
 Object &  Redshift & $M_{BH} (\times 10^9 M_\odot)$\\
 \hline
J002429.77+391319.0  & $6.620 \pm 0.004$ & $0.27\pm0.02$\\
J003836.10-152723.6  &$6.999 \pm 0.001$ & $1.36\pm0.05$\\
J004533.57+090156.9 &$6.441\pm 0.004$& $0.63\pm0.02$\\
J021847.04+000715.2 & $6.766\pm0.004$& $0.61\pm0.07$\\
J024655.90-521949.9 & $6.86\pm 0.02$& $1.05\pm 0.37$\\
J025216.64-050331.8  & $6.99\pm0.02$ & $1.28\pm 0.09$\\
J031343.84-180636.4 & $7.611 \pm 0.004$& $1.61\pm 0.40$\\
J031941.66-100846.0 & $6.816\pm0.004$ & $0.40\pm 0.03$\\
J041128.63-090749.8  & $6.827\pm0.006$ & $0.95\pm 0.09$\\
J043947.08+163415.7  & $6.519\pm0.003$& $0.63\pm 0.02$\\
J052559.68-240623.0 &$6.543\pm 002$& $0.002$ $0.29\pm $\\
J070626.39+292105.5  &$6.5925\pm 0.0004$& $2.11\pm 0.04$\\
J082931.97+411740.4 &$6.384\pm0.004$& $1.40 \pm 0.16$\\
J083737.84+492900.4 & $6.773\pm 0.007$& $0.71 \pm 0.18$\\
J083946.88+390011.5 &$6.702 \pm 0.001$& $0.81 \pm 0.02$\\
J091054.53-041406.8  & $6.9046\pm0.0003$& $0.671 \pm 0.003$\\
J092120.56+000722.9& $6.610\pm0.003$ & $0.41 \pm 0.03$\\
J092347.12+040254.4& $6.719\pm0.005$& $0.26 \pm 0.01$\\
J092359.00+075349.1&$6.5654\pm0.0002$ & $1.77 \pm 0.02$\\
J100758.26+211529.2&$6.682\pm0.002$ & $0.49 \pm 0.15$\\
J105807.72+293041.7 & $7.48\pm0.01$  & $1.43 \pm 0.22$\\
J110421.59+213428.8& $6.585\pm0.005$  & $0.54 \pm 0.03$\\
J112001.48+064124.3 & $6.766\pm0.005$  & $1.69 \pm 0.15$\\
J112925.34+184624.2  &$7.070\pm0.003$& $1.35 \pm 0.04$\\
J113508.93+501133.0& $6.824\pm0.001$& $0.29 \pm 0.02$\\
J121627.58+451910.7 &$6.579\pm0.001$&$1.49 \pm 0.05$\\
J131608.14+102832.8 &$6.648\pm0.003$&$0.61 \pm 0.20$\\
J134208.10+092838.6 &$7.51\pm0.01$&$0.81 \pm 0.18$\\
J153532.87+194320.1  &$6.370\pm0.001$&$3.53 \pm 0.33$\\
J172408.74+190143.0&$6.480\pm0.001$&$0.67 \pm 0.08$\\
J200241.59-301321.7&$6.673\pm0.001$&$1.62 \pm 0.27$\\
J210219.22-145854.0&$6.652\pm0.003$& $0.74 \pm 0.11$\\
J221100.60-632055.8 &$6.83\pm0.01$& $0.55 \pm 0.24$\\
J223255.15+293032.0  &$6.655\pm0.003$ &$3.06 \pm 0.36$\\
J233807.03+214358.2 &$6.565\pm0.009$& $0.56 \pm 0.03$\\
\hline
\end{tabularx}
		\caption{ 
		This table contains 35 high-redshift quasars at $z>6$ with their central SMBH mass from \cite{Yang2021}.}
			\label{table:6}
	\end{center}
\end{table*}

\end{document}